\documentclass[11pt]{article}
\usepackage{epsf}
\usepackage{axodraw}

\newcommand{ \centeron }[2]{{\setbox0=\hbox{#1}\setbox1=\hbox{#2}\ifdim
                             \wd1>\wd0\kern.5\wd1\kern-.5\wd0\fi \copy0
                             \kern-.5\wd0\kern-.5\wd1\copy1\ifdim\wd0>\wd1
                             \kern.5\wd0\kern-.5\wd1\fi}}
\newcommand{ \ltap }{\>\centeron{\raise.35ex\hbox{$<$}}
                     {\lower.65ex\hbox{$\sim$}}\>}
\newcommand{ \gtap }{\>\centeron{\raise.35ex\hbox{$>$}}
                     {\lower.65ex\hbox{$\sim$}}\>}

\newcommand{ \slashchar }[1]{\setbox0=\hbox{$#1$}   
   \dimen0=\wd0                                     
   \setbox1=\hbox{/} \dimen1=\wd1                   
   \ifdim\dimen0>\dimen1                            
      \rlap{\hbox to \dimen0{\hfil/\hfil}}          
      #1                                            
   \else                                            
      \rlap{\hbox to \dimen1{\hfil$#1$\hfil}}       
      /                                             
   \fi}                                             %

\newcommand{ \Mmess    }{M}
\newcommand{ \ra       }{\rightarrow}
\newcommand{ \textfrac }[2]{ {\textstyle\frac{#1}{#2}} }
\newcommand{ \ph       }{\gamma}

\def\singleandabitspaced{\baselineskip=\normalbaselineskip\multiply
    \baselineskip by 110\divide\baselineskip by 100}
\def\abstractspacing{\baselineskip=\normalbaselineskip\multiply
    \baselineskip by 110\divide\baselineskip by 100}
\def\singlespaced{\baselineskip=\normalbaselineskip}

\newcommand{\Journal}[4]{{#1}\ \textbf{#2}, #3 (#4)}  
\newcommand{ \NPB    }[3]{\Journal{Nucl. Phys.}{B#1}{#2}{#3}}
\newcommand{ \PLB    }[3]{\Journal{Phys. Lett. B}{#1}{#2}{#3}}

\newcommand{ \PRD    }[3]{\Journal{Phys. Rev. D}{#1}{#2}{#3}}

\newcommand{ \PRL    }[3]{\Journal{Phys. Rev. Lett.}{#1}{#2}{#3}}

\newcommand{ \PREP   }[3]{\Journal{Phys. Rep.}{#1}{#2}{#3}}

\newcommand{ \JHEP   }[3]{JHEP #1:#2 (#3)}

\newcommand{ \xxx    }[1]{\texttt{[#1]}}

\begin{document}

\singlespaced

\begin{titlepage}

\begin{flushright}
hep-ph/9909376 \\
September 1999 
\end{flushright}

\vspace{1.5cm}

\begin{center}
\mbox{\Large \textbf{Distinguishing anomaly mediation from gauge mediation}} \\
\vspace*{0.3cm}
\mbox{\Large \textbf{with a Wino NLSP}} \\

\vspace*{2.0cm}
{\Large Graham D. Kribs\footnote{\texttt{kribs@cmu.edu}}} \\
\vspace*{0.5cm}
\textit{Department of Physics, Carnegie Mellon University, 
        Pittsburgh, PA~~15213-3890} \\

\vspace*{1.0cm}

\begin{abstract}
\indent

\abstractspacing

A striking consequence of supersymmetry breaking communicated purely 
via the superconformal anomaly is that the gaugino masses are 
proportional to the gauge $\beta$-functions.  This result, however, 
is not unique to anomaly mediation.  We present examples of ``generalized'' 
gauge-mediated models with messengers in standard model representations 
that give nearly identical predictions for the gaugino masses,
but positive (mass)$^2$ for all sleptons.  There are remarkable
similarities between an anomaly-mediated model with a small 
additional universal mass added to all scalars and the gauge-mediated
models with a long-lived Wino next-to-lightest supersymmetric particle 
(NLSP), leading to only a small set of observables that provide
robust distinguishing criteria.  These include ratios of the
heaviest to lightest selectrons, smuons, and stops.  The sign of
the gluino soft mass an unambiguous distinction, but requires 
measuring a difficult class of one-loop radiative corrections
to sparticle interactions.  A high precision measurement of the
Higgs-$b$-$\overline{b}$ coupling is probably the most promising 
interaction from which this sign might be extracted.

\end{abstract}

\end{center}
\end{titlepage}

\newpage
\setcounter{page}{2}
\renewcommand{\thefootnote}{\arabic{footnote}}
\setcounter{footnote}{0}
\singleandabitspaced

\section{Introduction}
\label{introduction-sec}

Supersymmetry breaking communicated dominantly via the superconformal 
anomaly is a very interesting new approach to weak scale 
supersymmetry \cite{RS, GLMR}.  In the absence of singlets, 
anomaly mediation provides a one-loop contribution to the 
gaugino masses, a one-loop contribution to the trilinear scalar couplings, 
and a two-loop contribution to the scalar (mass)$^2$ of the minimal 
supersymmetric standard model (MSSM)\@.  These contributions
can be understood as arising from a super-Weyl invariant action 
once supersymmetry breaking is explicitly included in the 
superconformal compensator in supergravity, and are thus
precisely proportional to the gravitino mass.  
If there are no direct couplings between the MSSM sector 
and the supersymmetry breaking sector, then anomaly mediation 
provides the dominant contribution to all the MSSM fields.  
This is a natural expectation if the MSSM fields and the 
supersymmetry breaking sector fields are physically separated 
on different branes \cite{RS}.

There are several advantages to the anomaly mediation approach.  
The supersymmetric flavor problem is ameliorated, since the 
potentially dangerous contributions to off-diagonal squark (mass)$^2$ 
are suppressed.  The form of the expressions 
for the masses induced via the superconformal anomaly are exact to all 
orders \cite{GLMR, JJanom}, and determined by infrared physics,
namely the low energy $\beta$-functions.  Finally, the ratio of 
gaugino masses to scalar masses is order one (i.e.\ \emph{not} 
one-loop suppressed).

Gauge mediation \cite{DNS, GRreport} 
shares several features with anomaly mediation,
namely the supersymmetric flavor problem is also ameliorated, 
and masses are induced at one-loop for gauginos and (mass)$^2$ are 
induced at two-loops for scalars.  Furthermore, gauge-mediated
gaugino masses are (at leading order) proportional 
to the gauge (coupling)$^2$, identical to anomaly mediation.
The key phenomenological difference is that gauge-mediated
soft masses depend on the content of the messenger sector, 
whereas anomaly-mediated soft masses depend on the low energy 
$\beta$-function coefficients.  More generally, supersymmetry 
breaking masses are determined by ultraviolet physics in gauge mediation, 
and by infrared physics in pure\footnote{``Pure'' meaning an 
anomaly-mediated MSSM without any additions or modifications, 
and thus without a solution to the negative slepton (mass)$^2$ 
problem.} anomaly mediation.  In general the phenomenology is
expected to be quite different, but there is no a priori reason
why these two fundamentally different origins of supersymmetry
breaking masses could not be ``accidentally'' rather similar.
We show that a simple choice of messenger matter using standard model 
(SM) representations gives the 
same numerical result for the size of gaugino masses at leading order.  
Other messenger sectors that give similar results are also briefly
mentioned.  At next-to-leading order there are two-loop
contributions to gaugino masses that do not respect the leading-order
equivalence.  However, there is still a restricted range of gauge-mediated
parameters that give gaugino masses that are nearly equivalent to 
the next-to-leading order predictions of anomaly mediation.

The scalar spectrum of a gauge-mediated model is well defined once
the messenger sector is fixed.  This suggests that a gauge-mediated
model could be falsified simply by measuring several slepton and 
squark masses in addition to the gaugino masses, and then determining 
if the spectrum is self-consistent.  On closer inspection, however, 
the situation is not quite so trivial.  The first and second generation 
squark masses are nearly identical to the pure anomaly-mediated result
in ``generalized messenger'' gauge-mediated models.  (Third generation 
squark masses are somewhat more distinct, but complicated by 
left-right mixing.)
Furthermore, pure anomaly mediation predicts slepton (mass)$^2$
that are negative, requiring one of several proposed remedies
\cite{RS, GLMR, PR, GGW, Lutygroup, KSS, FM, KaplanKribs}. 
Each ``solution'' to the negative slepton (mass)$^2$ problem
must at least provide additional contributions to the slepton 
masses, and can have varying effects on the remainder of the mass
mass spectrum.  In this paper, we employ the simple phenomenological 
solution that merely adds a universal mass term to all of the 
scalar masses, leaving the gaugino masses unchanged \cite{GGW, FM}.
A more drastic alternative that, for example, shifts the gaugino 
masses from their anomaly-mediated values would be trivially distinct 
from the gauge-mediated models discussed here, and thus
need not be considered further.

There are at least two other general distinctions between gauge mediation
and anomaly mediation:  The Wino NLSP is not stable in gauge mediation,
and the sign of the gluino soft mass is opposite (negative) in
anomaly mediation.  The extent to which these distinctions are 
phenomenologically useful criteria is also discussed.

Our main purpose in this paper is not to advocate that the gauge-mediated 
Wino NLSP models are more (or less) favored than an anomaly-mediated
model with an appropriate negative slepton (mass)$^2$ solution.
Instead, we are interested in determining the ways to experimentally verify
(or falsify) these scenarios.  
Our starting point is that a gaugino mass spectrum that is approximately
proportional to gauge $\beta$-functions is \emph{not} sufficient
to identify anomaly mediation as the source of supersymmetry breaking.  
Instead, several other criteria must be used to separate the 
gauge-mediated models discussed here from anomaly mediation.  
In this way we attempt to gain a more robust understanding of 
the signals of both anomaly mediation and gauge mediation.

\section{Constructing a model}
\label{constructing-sec}

In the following, we consider two complete different supersymmetric
models, each with fundamentally different means by which supersymmetry
breaking is communicated to the MSSM\@.  First, consider a model
with no (hidden sector) singlets such that anomaly mediation (AM) 
provides the dominant contribution to the gaugino masses.  In this
case, the expressions for the gaugino masses are \cite{RS, GLMR}
\begin{eqnarray}
M_a^{\rm AM} &=& \frac{\beta_a}{g_a} m_{3/2}^{\rm AM} \nonumber \\
    &\simeq& B_a^{(1)} \frac{g_a^2}{16 \pi^2} m_{3/2}^{\rm AM} \; ,
\label{AM-gaugino-eq}
\end{eqnarray}
where $m_{3/2}^{\rm AM}$ is the gravitino 
mass,\footnote{To avoid confusion, $m_{3/2}^{\rm AM}$ 
($m_{3/2}^{\rm GM}$) always corresponds to the value of the 
gravitino mass in the anomaly mediation (gauge mediation) model.}
$g_a$ is the gauge 
coupling,\footnote{$g_1$ is always taken to be in the GUT 
normalization, $g_1 = \sqrt{5/3} g'$.} and $B_a^{(1)} = (33/5, 1, -3)$ 
correspond to the one-loop $\beta$-function coefficients for
$a=[$U(1)$_Y$, SU(2)$_L$, SU(3)$_c]$.  For the purposes of this
section, only effects to leading order (i.e.\ to one-loop for gaugino 
masses) will be discussed.  The scalar masses in this model 
are generated both by anomaly mediation as well as an unspecified
source that provides an additional universal mass sufficient
to cure the slepton mass problem.  The general form is \cite{RS, 
GLMR, GGW}
\begin{eqnarray}
\tilde{m_i}^2 &=& -\frac{1}{4} \left[ 
    \beta_a \frac{\partial \gamma_i}{\partial g_a} 
  + \beta_{Y} \frac{\partial \gamma_i}{\partial Y} \right] 
    \, (m_{3/2}^{\rm AM})^2
  + m_0^2
\end{eqnarray}
where $\gamma_i$ is the anomalous dimension of the $i$ chiral 
superfield.  A sum over gauge couplings and Yukawa couplings is 
implicit.  Trilinear scalar couplings are also generated at one-loop,
and expressions can be found in \cite{RS, GLMR}.

These expressions for the soft masses induced by anomaly mediation 
are in general quite different from minimal messenger 
gauge mediation models and ordinary supergravity models.
There is, however, the apparent similarity that the 
anomaly mediation model generates gaugino masses at one-loop 
and scalar (mass)$^2$ at two-loops in a manner analogous to 
gauge mediation.  In addition, the one-loop expression for 
the gaugino mass depends in both models on $g^2/16\pi^2$, although 
the coefficients are different and the supersymmetry breaking mass 
is distinct.

Instead of restricting to messengers in complete GUT 
representations \cite{DNS}, consider generalizing the messenger 
sector of a gauge mediation model to a sum over arbitrary vector-like 
representations \cite{MartinGMSB}.\footnote{See also 
Refs.~\cite{DTW, CDM} for examples of models with gaugino
masses not in the canonical proportions.}
Supersymmetry breaking is present with a non-zero vev for 
the auxiliary components of the 
messenger fields, and after integrating them out, they induce the 
following gaugino masses through gauge mediation (GM)
\begin{eqnarray}
M_a^{\rm GM} &=& 
    \frac{g_a^2}{16 \pi^2} \sum_i S_a(i) g(F_i/M_i^2) \frac{F_i}{M_i} \; .
\label{GM-gaugino-general-eq}
\end{eqnarray}
The sum is over all messengers labeled by $i$, $S_a(i)$ is the 
Dynkin index for the $a$ gauge group, and $F_i$ and $M_i$ are the 
$F$-terms and fermion masses of the messengers.  The function $g(x)$ is
\begin{eqnarray}
g(x) &=& \frac{1}{x^2} \left[ (1+x) \log (1+x) + (1-x) \log (1-x) \right] \; ,
\label{g-eq}
\end{eqnarray}
and is equal to about $1, 1.05, 1.22, 1.39$ for $x = 0, 0.5, 0.9, 1$.  
In the approximation $F_i = F \ll M_i^2 = M^2$, meaning that 
the messengers have approximately the same supersymmetric and
supersymmetry breaking vevs, the sum is only over 
the Dynkin indices of the messengers $n_a \equiv \sum_i S_a(i)$, 
and then Eq.~(\ref{GM-gaugino-general-eq}) simplifies to
\begin{eqnarray}
M_a^{\rm GM} &\simeq& n_a \frac{g_a^2}{16 \pi^2} \frac{F}{M} \; .
\label{GM-gaugino-eq}
\end{eqnarray}
Note the striking formal similarity between the expression for the 
gaugino mass in the anomaly mediation model, 
Eq.~(\ref{AM-gaugino-eq}), and the expression for the gaugino
mass in the gauge mediation model, 
Eq.~(\ref{GM-gaugino-eq}).  Both are characterized by a discrete 
quantity multiplied by $g^2/16\pi^2$ multiplied by a supersymmetry 
breaking mass.  It is precisely this similarity that we 
exploit in the following to construct a gauge mediation model with gaugino
masses that are equivalent to anomaly mediation, using an appropriate
relationship between the supersymmetry breaking masses.  Note that
the gaugino mass and gauge coupling in Eq.~(\ref{AM-gaugino-eq}) 
are evaluated at the weak scale, whereas the gaugino mass and 
gauge coupling in Eq.~(\ref{GM-gaugino-eq}) are evaluated at 
the messenger scale.  The latter expression does not, however, acquire
a renormalization group correction to the order we are working 
since $M/g^2$ is one-loop renormalization group invariant.

\subsection{Generalized messenger models}

Utilizing the above generalization of the messenger sector, 
we now proceed to construct a gauge-mediated model with gaugino 
masses proportional to the one-loop gauge $\beta$-functions.
Require
\begin{eqnarray}
n_a &=& |B^{(1)}_a|
\label{relate-eq}
\end{eqnarray}
and that the supersymmetry breaking mass parameters coincide,
$F/M = m_{3/2}^{\rm AM}$.  The predictions for the
gaugino masses are identical to those of the anomaly mediation 
model, up to the sign of the gluino soft 
mass.\footnote{If a higher rank group associated
with the messengers broke to SU(3)$_c$, it is possible that the 
effective $n_3$ could be negative due to gauge messengers 
\cite{GiudiceRattazziwf}.  However, this also causes the 
(mass)$^2$ for at least the first and second generation squarks 
to be negative, and therefore does not appear to be viable.}
We emphasize that this is an accidental equivalence between
two completely separate origins of supersymmetry breaking with 
$m_{3/2}^{\rm GM} \ll m_{3/2}^{\rm AM}$.
The equivalence does not apply to the sign of gluino soft mass,
but this difference has only a limited phenomenological impact
on the spectra.  The sign does affect the gaugino 
soft mass predictions at next-to-leading (two-loop) order, and we 
discuss this in more detail in Sec.~\ref{higher-order-sec}.  
In principle, measuring this sign would be an unambiguous way of 
distinguishing these models, but experimentally this is rather 
difficult, as we explain in Sec.~\ref{gluino-sign-sec}.
Suffice to say there is no easily measurable difference between a 
model with a positive gluino soft mass, such as gauge mediation, 
and a model with a negative gluino soft mass, such as anomaly mediation.

The ratio $F/M$ that sets the overall scale of the gauge-mediated 
soft masses could be different from $m_{3/2}^{\rm AM}$
while the gaugino masses remain equivalent.
At first glance, only the proportionality ($n_1$ : $n_2$ : 
$n_3$) $=$ ($B^{(1)}_1$ : $B^{(1)}_2$ : $|B^{(1)}_3|$) is relevant.  
However, restricting to a set of messengers that preserves the 
perturbativity (but not necessarily the equivalence) of the gauge 
couplings up to the purported unification scale $\sim 10^{16}$~GeV 
implies that $F/M$ cannot be an integer multiple of 
$m_{3/2}^{\rm AM}$ (other than unity).  
Possible fractional values (such as 
$\textfrac{1}{2} m_{3/2}^{\rm AM}$ or 
$\textfrac{3}{2} m_{3/2}^{\rm AM}$) 
could only occur with messengers in non-vectorlike multiplets,
that can be justifiably ignored due to the difficulty of giving such 
fields a large supersymmetric mass.  Under these constraints,
Eq.~(\ref{relate-eq}) can be expanded
\begin{eqnarray}
\textfrac{1}{5} \left( n_Q + 8 n_u + 2 n_d + 3 n_L + 6 n_e \right) &=& 
    \textfrac{33}{5} \nonumber \\
3 n_Q + n_L &=& 1 \\
2 n_Q + n_u + n_d &=& 3 \nonumber \; ,
\end{eqnarray}
where $n_X$ corresponds to the number of $X + \overline{X}$ pairs
of vectorlike messenger multiplets in the SM representations
($Q$, $u$, $d$, $L$, $e$).  The set of solutions are 
characterized by
\begin{eqnarray}
n_Q \; = \; 0 \quad & & \quad n_L \; = \; 1 \nonumber \\
n_u + n_d &=& 3 \label{mess-content-eq} \\
n_u + n_e &=& 4 \; , \nonumber
\end{eqnarray}
meaning ($n_Q$, $n_u$, $n_d$, $n_L$, $n_e$) can only be one of 
($0$, $0$, $3$, $1$, $4$), 
($0$, $1$, $2$, $1$, $3$), 
($0$, $2$, $1$, $1$, $2$), or 
($0$, $3$, $0$, $1$, $1$).  These sets of multiplets are degenerate 
at leading order, but give slightly differing results at next-to-leading
order (e.g.\ two-loop expressions for gaugino masses, 
and two-loop contributions to the gauge $\beta$-functions above 
the messenger scale).  As as aside, we note that the third set of 
multiplets corresponds to one $5 + \overline{5}$ and two nearly 
complete $10 + \overline{10}$'s, but it is conspicuously missing the 
two pairs of $Q + \overline{Q}$'s.  This may provide a useful starting 
point for a dynamical determination of the above sets of messenger
fields, although this is beyond the scope of this paper.  
In any case, determining the sets of multiplets 
that realize the relation Eq.~(\ref{relate-eq}), and in particular
that only SM representations are needed, is one of the important 
results of this paper.

\subsection{Multi-singlet models}

Another approach to constructing a gauge mediation model is to expand
the messenger sector such that there are several singlets (see
also Ref.~\cite{Wagner}) with
either different supersymmetric masses, or different $F$-terms,
or both.  This approach has the advantage that matter in complete SU(5) 
representations is sufficient, thus naively preserving one-loop 
gauge-coupling unification.  However, the supersymmetric mass scales
or the $F$-terms (or both) must differ among the SU(3) $\times$ SU(2) 
$\times$ U(1) components fields, breaking the SU(5) ansatz.  

There are two potential benefits of modifying the supersymmetric mass 
scale of SU(3) $\times$ SU(2) $\times$ U(1) component messenger fields.
The first is a threshold effect, Eq.~(\ref{g-eq}), whose
size depends on $F/M^2$.  A given gaugino mass could be increased
(relative to $M \ra$ large, with fixed $F$) by at most about 
$35\%$, if $M$ is rather close to $\sqrt{F}$.  In general it is hard 
to imagine how this could arise dynamically, although some ideas
have been discussed in Ref.~\cite{KSS} (in an anomaly mediation context).
The second potential benefit of shifting the supersymmetric mass scale 
of messengers exploits the running gauge coupling.  The gaugino
mass induced at the messenger scale is proportional to the gauge
coupling squared evaluated at the messenger scale $g^2(M)$, and so 
it is possible to shift a given gaugino mass by a factor
$g^2(M_{\rm new})/g^2(M_{\rm old})$.  
This effect, however, is really a next-to-leading order correction.
In practice, only $g_3^2$ evolves significantly (by at most about $40\%$)
between the lowest and highest messenger scales (between about $10^5$ 
to $10^9$ GeV) that are consistent with gauge mediation giving the 
dominant contribution to soft masses.  One difficulty with both
of these approaches is that several SM component fields of (say) 
complete SU(5) reps are charged under more than one gauge group of
the SM, so that modifying the scale of a given pair of messengers 
affects several gaugino masses simultaneously.  We conclude that 
effects resulting from shifting the supersymmetric mass scale of
messengers, by themselves, cannot reproduce any of the large ratios 
$B_1^{(1)}/|B_3^{(1)}| = 11/5$ or $|B_3^{(1)}|/B_2^{(1)} = 3$.

If several singlets communicate supersymmetry breaking from the
dynamical supersymmetry breaking (DSB) sector to the messengers, 
it is not implausible that they could
couple differently to SU(3) $\times$ SU(2) $\times$ U(1) component 
messenger fields.  Then, given a small hierarchy of $F$-terms,
is it trivial to construct a gauge-mediated model that has
gaugino masses in a proportion indistinguishable from
anomaly mediation.  For example, take the components of a 
10 + $\overline{10}$ coupled to two SM singlets $X_1$ and $X_2$
using the messenger superpotential
\begin{eqnarray}
W &=& X_1 Q \overline{Q} + X_2 u \overline{u} + X_2 e \overline{e} \; .
\end{eqnarray}
With $F_{X_1}/(3 M_{X_1}) = 7 F_{X_2}/(3 M_{X_2}) = 
m_{3/2}^{\rm AM}$, this model 
generates gaugino masses in exactly the same proportion as the one-loop 
$\beta$-function coefficients.\footnote{It is also possible that 
$F$-terms of different singlets could have opposite signs and be
arranged such that the gaugino masses are in the same proportion
as anomaly mediation including the sign of $M_3$.  However, 
we are not aware of any DSB or messenger model that could give
this result.}

\subsection{Properties of the models}

The two classes of models discussed above, namely the generalized
messenger models and the multi-singlet model, generate the same result 
for the gaugino masses, but give somewhat different results for
the scalar (mass)$^2$:
\begin{eqnarray}
m_i^2 &=& 2 \frac{F^2}{M^2} \sum_a C_a(i) \frac{g_a^4}{(16 \pi^2)^2} n_a 
   \qquad\qquad\qquad\qquad\qquad\, \mbox{(generalized messengers)} 
   \label{gm-scalar-eq} \\
m_i^2 &=& 2 \frac{F^2}{M^2} \sum_a C_a(i) \frac{g_a^4}{(16 \pi^2)^2} \left[ 
   m_a(X_1) \frac{F_{X_1}^2}{F^2} + m_a(X_2) \frac{F_{X_2}^2}{F^2} \right] \; .
   \qquad \mbox{(multi-singlet)} \label{ms-scalar-eq}
\end{eqnarray}
The $m$ factors are $m_a(X_1) = (1/5, 3, 2)$ and $m_a(X_2) = (14/5, 0, 1)$ 
respecting $m_a(X_1) + m_a(X_2) = (3, 3, 3)$, and we have taken
the mass scale of the messengers $M$ to be the same for both models.
Notice that each gauge group is weighted by $n_a$ in the generalized 
messenger models, whereas each gauge group is effectively weighted 
by $n_a^2$ in the multi-singlet model since each scalar (mass)$^2$ 
in the latter is proportional to $F^2$.
Thus, holding the gaugino mass spectrum fixed, these two classes of 
models give different predictions for the scalar masses.  For example,
squark masses tend to be about $15\%$ lighter in the generalized messenger
models.  This illustrates that without specifying a particular messenger 
model, there is no unique prescription to translate a gaugino mass 
spectrum into a scalar mass spectrum.

Gauge coupling unification does not occur at $10^{16}$ GeV for the 
generalized messenger models; instead, $g_1$ is typically much larger 
than $g_3$, which is somewhat larger than $g_2$.  We verified that 
$g_1$ remains perturbative near $10^{16}$ GeV when calculated 
to two-loop order, as long as the messenger scale is larger 
than about $10^6$ GeV\@.  Ordinarily
the unification scale is defined by $g_1 \simeq g_2$, which
in the generalized messenger models occurs at an intermediate 
scale $\sim 10^{10} \ra 10^{12}$ GeV\@.  Conversely, gauge coupling
unification can occur in the multi-singlet model as long as the
shifts in the $\beta$-functions due to the additional messenger
fields are nearly independent of the gauge group (such as fields 
filling complete SU(5) reps).

A fascinating property of the generalized messenger models is that
the predictions for the first and second generation squark masses 
are nearly identical to pure anomaly mediation.  This is evident from
Eq.~(\ref{gm-scalar-eq}) at the messenger scale, where the dominant 
contribution proportional to $g_3^4$ is the same as anomaly mediation.
The prediction is also very well preserved under renormalization group 
(RG) evolution between the messenger scale and the weak scale, since the 
first and second generation scalar mass relations induced by 
gauge mediation are very close to the renormalization group 
invariant mass relations of anomaly mediation.  
This coincidence occurs precisely because $n_3$ is opposite in sign 
to $B_3^{(1)}$.  Hence, this does not occur for slepton masses.
Due to this interesting property, we concentrate most of 
the remaining discussion of gauge-mediated Wino NLSP models on the 
generalized messenger models.

The mostly Wino lightest neutralino is the NLSP and typically decays
into a gravitino and a photon, although heavier NLSPs can also 
have a significant branching fraction into a gravitino and a $Z$ 
\cite{AKKMM, BMPZ, CDM}.  Depending on 
whether the fundamental supersymmetry breaking scale is smaller or 
larger than about a few hundred TeV, the Wino NLSP could decay 
either inside or outside a typical collider detector.
If the decay NLSP $\ra$ gravitino $+$ photon were to 
occur well within a detector, it would be
clearly evident with a hard photon emitted for every NLSP produced
either directly or indirectly.  This is a very robust signal in 
gauge mediation \cite{AKKMM} and completely different from
anomaly mediation.  However, if the decay length is significantly
longer than a typical collider detector,\footnote{The decay length
scales as the fourth power of the fundamental supersymmetry
breaking scale, and therefore could be anywhere from microns to the 
distance from the Earth to the Sun.} the long-lived Wino NLSP 
is indistinguishable from a stable Wino LSP\@.  We concentrate
on this scenario for the remainder of the paper.

In supersymmetric models with a Wino (N)LSP, the mass splitting between 
the lightest chargino $\tilde{C}_1^\pm$ and the lightest neutralino 
$\tilde{N}_1$ is very small since both fields are nearly pure 
Wino-like states.  Expressions for the mass splitting at 
tree-level \cite{CDM, GGW} and at one-loop \cite{PBMZ, CDM} have been 
calculated, with the intriguing possibility of a macroscopic 
$\tilde{C}_1^\pm \ra \tilde{N}_1 \pi^\pm$ decay length signal that 
has been studied in detail in Refs.~\cite{macroscopicSUGRA, FMRSS-GM}.
This is also an interesting signal of the gauge-mediated Wino NLSP
models discussed here.  However, it is not a useful distinction
between gauge mediation and anomaly mediation because the decay
lengths of the respective Wino (N)LSPs are comparable, as 
discussed below.

\begin{figure}[!t]
\centering
\epsfxsize=5.0in
\hspace*{0in}
\epsffile{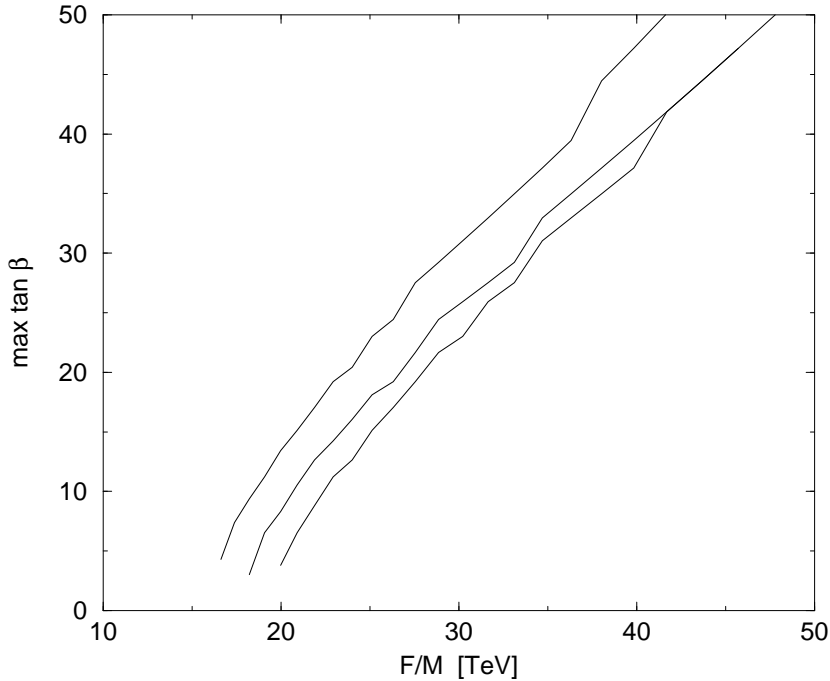}
\caption{Maximum $\tan\beta$ as a function of $F/M$ in generalized
messenger models by requiring that $m_{\tilde{\tau}_1}$ is larger
than the current LEP bound.  The bottom, middle, and top lines 
correspond to $M = 10^5$, $10^7$, and $10^9$ TeV\@.  The parameter
space above and to the left of the lines is excluded.}
\label{tbeta-fig}
\end{figure}

A complete set of parameters characterizing these models must also
include $\tan\beta$ and $\mu$.
Demanding the proper electroweak symmetry breaking vacuum with the 
correct value of $M_Z$ determines $\mu^2$ as a function of the Higgs 
soft masses and $\tan\beta$ (at tree-level), leaving only
the sign of $\mu$ unknown.  Requiring that $m_{\tilde{\tau}_1}$ 
be greater than the current LEP bound (of about $90$ GeV)
implies that there is an upper bound on $\tan\beta$ as a function
of the messenger parameters, shown in Fig.~\ref{tbeta-fig}.
Above about $F/M = 50$ TeV the limit disappears because the
lightest stau mass is always larger than the LEP bound.

\section{Distinctions between anomaly mediation and gauge mediation}
\label{distinctions-sec}

There are three central phenomenological differences between
anomaly mediation and gauge mediation: the scalar spectrum
is in general different, the sign of the gluino soft mass
is different, and the Wino NLSP of gauge mediation is unstable.
However, none of these differences are necessarily trivial to 
establish in a collider experiment, as discussed below.

\subsection{Scalar spectrum}
\label{scalar-distinction-sec}

\begin{figure}[!t]
\centering
\epsfxsize=5.0in
\hspace*{0in}
\epsffile{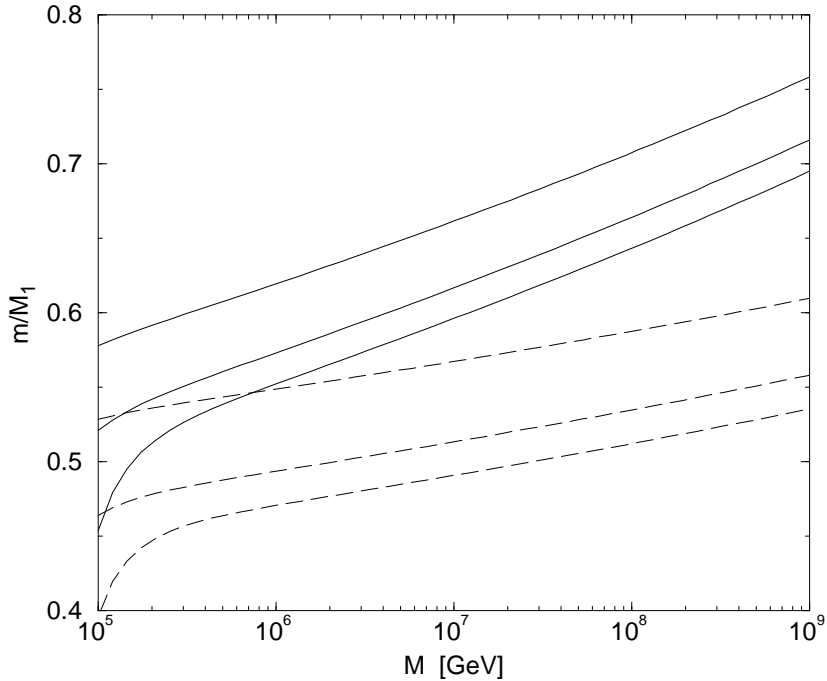}
\caption{Contours of $m_{\tilde{e}_R}/M_1$ (solid lines) and 
$m_{\tilde{e}_L}/M_1$ (dashed lines) as a function of the messenger 
scale $\Mmess$.  The top, middle, and bottom lines correspond to 
$F/M = 20$, $40$, and $80$ TeV, respectively.}
\label{eLeRoverM1-fig}
\end{figure}

Once the overall supersymmetry breaking scale $F/M$ is established, the 
scalar spectrum of the gauge-mediated models is fixed, up to a logarithmic
sensitivity to the messenger scale.  Unlike anomaly mediation,
there is no reason to suggest there should be additional contributions
to the matter scalars (not including the Higgs scalars)
if gauge mediation via SM interactions provides the dominant source 
of supersymmetry breaking.  Indeed, the elegant resolution to 
the supersymmetric flavor problem via gauge mediation would, in general, 
be lost with additional contributions (unless they were 
flavor-independent, aligned, or very heavy\footnote{See 
Ref.~\cite{KaplanKribsflavor} for a well-motivated example of just 
such a possibility.}).  Thus, the simplest way to exclude the
gauge-mediated models discussed here is to experimentally verify that 
the scalar mass spectrum does \emph{not} follow the gauge-mediated
scalar spectrum.  For example, the charged selectron masses
fall in a relatively narrow mass range relative to 
$M_1$,\footnote{The ratio to the Bino mass (instead of the lighter
Wino mass) was chosen to minimize dependence on higher order 
corrections (see Sec.~\ref{higher-order-sec}).}
shown in Fig.~\ref{eLeRoverM1-fig}.
The size of the slepton masses in anomaly mediation, by contrast, 
are dependent on $m_0$, and therefore a priori completely unrelated 
to $M_1$.

\begin{figure}[!t]
\centering
\epsfxsize=5.0in
\hspace*{0in}
\epsffile{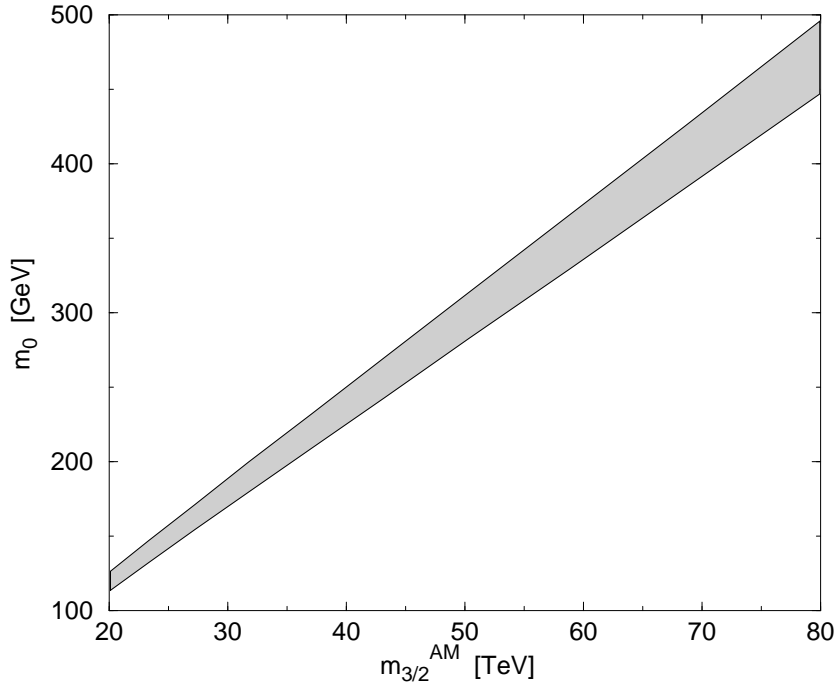}
\caption{Range of $m_0$ as a function of $m_{3/2}^{\rm AM} = F/M$ 
that implies the anomaly-mediated and generalized gauge-mediated models 
predict the same left-handed first and second generation slepton masses.  
The shaded band is the result of varying the messenger scale $\Mmess$ 
between $10^5$ to $10^9$ GeV in the gauge-mediated model.}
\label{m0-Lambda-fig}
\end{figure}

In anomaly mediation, it is not at all unreasonable that $m_0$ may be 
moderately small, 
meaning that it gives a significant contribution to sleptons to 
render them positive, but gives an insignificant contribution 
(or none at all) to squarks.  This of course depends on the 
underlying origin of $m_0$, of which we remain agnostic.  There is,
in fact, a range of $m_0$ that implies the anomaly-mediated 
and gauge-mediated predictions for the left-handed first and second 
generation slepton masses are identical.  The parameter space of
this ``worst nightmare'' situation is shown in Fig.~\ref{m0-Lambda-fig}.
Note that including experimental uncertainties would widen the 
overlapping range of $m_{3/2}^{\rm AM} = F/M$ for a given $m_0$.

It is therefore possible that the only differences in the 
scalar sector between an anomaly-mediated model with $m_0$ as 
shown in Fig.~\ref{m0-Lambda-fig} and a generalized gauge-mediated 
model are ultimately related to $m_{\tilde{\ell}_R}$, $\mu$, 
the trilinear scalar couplings, or the presence of Yukawa 
contributions to the third generation scalars in anomaly mediation.  
In the remainder of 
this section we discuss to what extent these observables
provide useful distinguishing criteria between anomaly mediation 
and gauge mediation.

In the generalized gauge-mediated Wino NLSP models, the first and 
second generation right-handed slepton mass $m_{\ell_R}$ is always 
larger than the corresponding 
left-handed slepton mass $m_{\ell_L}$, due to the large ratio
$n_1/n_2 = 33/5$.  Explicitly, there are three contributions to
the slepton mass difference $m_{\ell_R}^2 - m_{\ell_L}^2$:
the gauge-mediated contribution at the messenger scale, the RG 
contribution, and the $D$-term contribution.  The $D$-term
contribution is accidentally rather small due to a numerical 
cancellation, and can be neglected.  The other contributions are
\begin{eqnarray}
\left[ m_{\ell_R}^2 - m_{\ell_L}^2 \right]_{\rm mess} &=&
\frac{3}{2} \frac{n_2}{(16 \pi^2)^2} \left[ \frac{3 n_1}{5 n_2} g_1^4 
    - g_2^4 \right] \frac{F^2}{M^2} \\
\left[ m_{\ell_R}^2 - m_{\ell_L}^2 \right]_{\rm RG} &\simeq&
    6 \frac{n_2^2}{(16 \pi^2)^3} \left[ \frac{3 n_1^2}{5 n_2^2} g_1^6
    - g_2^6 \right] \frac{F^2}{M^2} \ln \frac{M}{m_\ell} \; , 
\end{eqnarray}
that in practice are numerically roughly comparable.  The ratio
can be approximately written as
\begin{eqnarray}
\frac{m_{\ell_R}^2 - m_{\ell_L}^2}{m_{\ell_L}^2} \simeq 
(0.25 \pm 0.05) + 0.04 \, \ln \frac{M}{10^5 \; \rm GeV} \; ,
\end{eqnarray}
where the $\pm 0.05$ arises from the variation of $F/M$
throughout the range $20 \ra 80$ TeV\@.  This is very different from 
anomaly mediation, where the difference between the left-handed and 
right-handed slepton masses is less than a few percent 
throughout most of the parameter space \cite{GGW}.

The bilinear supersymmetric Higgs mass, $\mu$, feeds into 
several low energy observables, including the heavier chargino, 
the two heaviest neutralinos (assuming $\mu$ is larger than $M_1$), 
the heavier Higgs scalar masses, and the off-diagonal left-right
(LR) squark and slepton mixing.  Hence, determining the value
of $\mu$ from experiment can be done via several different 
classes of signals.

In both the anomaly-mediated and gauge-mediated models discussed 
here, it is generally a good approximation throughout the parameter space 
that $\mu^2 \sim -m_{H_u}^2$.  In anomaly mediation, there is an 
interesting (apparently accidental) cancellation between the
renormalization group contributions that feed into $m_{H_u}^2$
due to the presence of a nonzero $m_0$ (that breaks the 
renormalization group invariance of the anomaly-mediated 
spectrum).  The result is a ``focusing'' effect \cite{FM, focus2} 
that renders $m_{H_u}^2$ quite insensitive to large changes in $m_0$.  
In particular, $m_{H_u}^2$ is determined essentially by just the 
anomaly-mediated value that is approximately
\begin{eqnarray}
m_{H_u}^2 &\simeq& Y_t^2 \left(- 16 g_3^2 - 9 g_2^2 + 18 Y_t^2 \right)
\left( \frac{m_{3/2}^{\rm AM}}{16 \pi^2} \right)^2 \; ,
\end{eqnarray}
or that roughly $(-m_{H_u}^2)^{1/2}$ is about 
$2.5 \, m_{3/2}^{\rm AM}/(16 \pi^2)$.
Therefore, in an anomaly-mediated model with a universal additional scalar
mass $m_0$, the value of $\mu$ is fixed once the scale of the 
gaugino masses has been established.

In gauge mediation the Higgs scalar masses are also determined
once the messenger sector is fixed and the scale of the gaugino
masses has been established.  However, no dynamical origin
for $\mu$ was given, and indeed it is possible that
the Higgs soft masses could be affected by the mechanism that
ultimately determines $\mu$ (see e.g.~\cite{mu-GMSB, DTW}).
For this reason, observables that depend on $\mu$ 
are not particularly reliable distinctions between anomaly mediation
and gauge mediation, unless the gauge-mediated contributions
to the Higgs soft masses dominate over all other possible
contributions.

Another interesting observable is the decay length of 
$\tilde{C}_1^\pm \ra \tilde{N}_1 \pi^\pm$.  This is also, 
unfortunately, not a useful distinction between anomaly mediation 
and gauge mediation for two reasons.  First, the
one-loop corrections dominate throughout the parameter
space of interest, and to a very good accuracy depend only on 
kinematical functions of $M_2$ and $M_W$ \cite{CDM}.  This
contribution is therefore the same for a given Wino mass.
Second, the smaller tree-level corrections \cite{CDM, GGW}
depend sensitively on $\mu$ (and $\tan\beta$), that we have
argued is not a reliable distinction.  Thus, while the 
macroscopic decay length of the lightest chargino is an excellent
signal of a Wino (N)LSP, it does not provide any useful information
to distinguish the mediation of supersymmetry breaking.

Finally, at leading order trilinear scalar couplings, $A_f$,
are generated in anomaly mediation, but not in 
gauge mediation.  They do reappear in gauge mediation after 
renormalization group evolution to the weak scale, but are 
usually smaller (in absolute value) than and opposite in sign 
to the anomaly-mediated values.  These couplings affect the
(mass)$^2$ matrix for the sfermions, but to a good approximation
for moderate to large $\tan\beta$, they only significantly 
impact the stop mass matrix.  This is because the
off-diagonal term for up-type sfermions is $m_f (A_f - \mu / \tan\beta)$
whereas for down-type sfermions it is $m_f (A_f - \mu \tan\beta)$, 
which shows that the term proportional to $\mu$ is significantly
diminished (enhanced) relative to $A_f$ for up-type (down-type)
sfermions.  Thus, the splitting between the heavier stop ($\tilde{t}_2$) 
and the lighter stop ($\tilde{t}_1$) mass eigenstates is generally 
larger in anomaly mediation.  Taking a ratio, such as 
$(m_{\tilde{t}_2}^2 - m_{\tilde{t}_1}^2)/m_{\tilde{t}_2}^2$, also
eliminates the dependence on $m_0$ in anomaly mediation,
and therefore provides a useful additional distinction.  Note, however, 
that the gauge mediation prediction of nearly zero trilinear scalar 
couplings at the messenger scale relies on the assumption that 
the couplings between messenger fields and MSSM matter fields are very 
small \cite{GiudiceRattazziwf}.  

Four example models' spectra are given in Appendix~\ref{example-models-app}
to illustrate the comparison between two anomaly mediation models
and two gauge mediation models.

\subsection{Sign of the gluino soft mass}
\label{gluino-sign-sec}

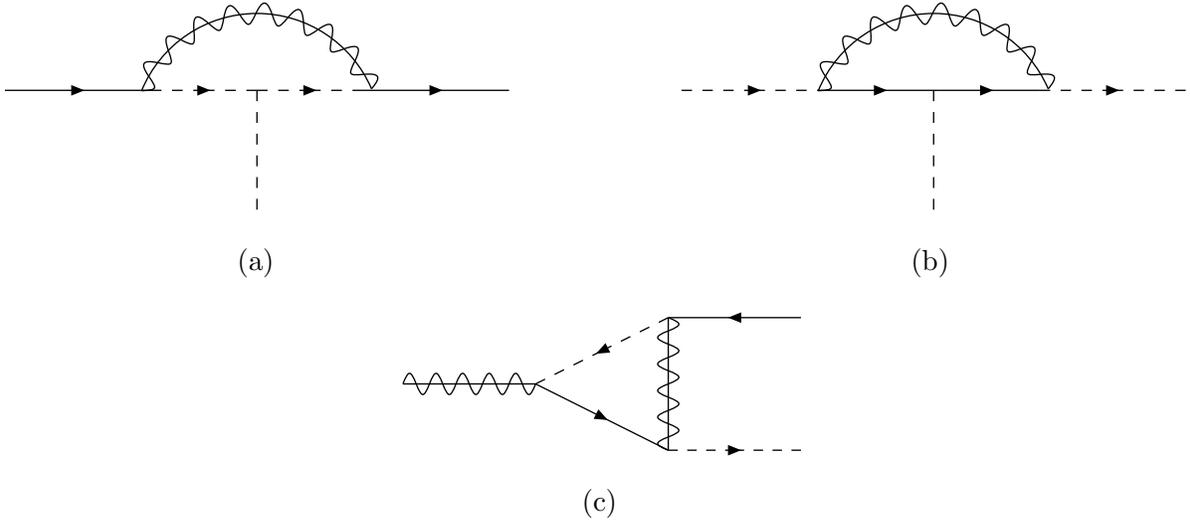
\begin{figure}[!t]
\begin{picture}(455,110)(0,0)
%
%
   \ArrowLine(     5, 70 )(  60, 70 )
   \DashArrowLine( 60, 70 )( 100, 70 ){4}
   \CArc(          100, 52 )(  47, 23.3, 157.7 )
   \PhotonArc(     100, 52 )(  47, 23.3, 157.7 ){4}{9}
   \DashArrowLine( 100, 70 )( 140, 70 ){4}
   \ArrowLine(     140, 70 )( 195, 70 )
   \DashLine(      100, 70 )( 100, 25 ){4}
   \Text(          100,  5 )[c]{(a)}
%
%
   \DashArrowLine( 260, 70 )( 315, 70 ){4}
   \ArrowLine(     315, 70 )( 355, 70 )
   \CArc(          355, 52 )(  47, 23.3, 157.7 )
   \PhotonArc(     355, 52 )(  47, 23.3, 157.7 ){4}{9}
   \ArrowLine(     355, 70 )( 395, 70 )
   \DashArrowLine( 395, 70 )( 450, 70 ){4}
   \DashLine(      355, 70 )( 355, 25 ){4}
   \Text(          355,  5 )[c]{(b)}
\end{picture}
\begin{picture}(455,90)(0,0)
%
%
   \Line(          155, 50 )( 205, 50 )
   \Photon(        155, 50 )( 205, 50 ){4}{5}
   \ArrowLine(     205, 50 )( 255, 25 )
   \DashArrowLine( 255, 25 )( 305, 25 ){4}
   \ArrowLine(     305, 75 )( 255, 75 )
   \DashArrowLine( 255, 75 )( 205, 50 ){4}
   \Line(          255, 25 )( 255, 75 )
   \Photon(        255, 25 )( 255, 75 ){4}{5}
   \Text(          230,  5 )[c]{(c)}
\end{picture}
\caption{One-loop diagrams that are proportional to the phase 
${\rm arg}(M_3)$ of the gluino soft mass: 
(a) a one-loop correction to a fermion-fermion-Higgs interaction,
(b) a one-loop correction to a trilinear scalar coupling, and 
(c) a one-loop correction to the gaugino-quark-squark coupling.
The arrows denote the flow of baryon number.}
\label{one-loop-fig}
\end{figure}

In general, the soft mass of the gluino in the MSSM may be complex.
After a field redefinition of the gluino, the phase of the gluino soft
mass appears only in the interaction of a gluino and a chiral multiplet
(see Appendix~\ref{gluino-app} for details).  
Furthermore, the phase cancels in 
processes that involve a vertex with a chiral multiplet \emph{and} 
its hermitian conjugate.  The phase reappears only in a chirality 
violating interaction (such as an interaction with a Higgs) or a 
``fermion number violating'' interaction.  One-loop examples of such
interactions are shown in Fig.~\ref{one-loop-fig}.

A significant CP-violating phase in the gluino soft mass, i.e.\ 
${\rm arg}(M_3)$ not close to $0$ or $\pi$, gives large 
contributions to CP-violating processes, particularly the electric 
dipole moment of the neutron \cite{EDM, IN98}.  
Here we are interested in the processes of 
Fig.~\ref{one-loop-fig} that can distinguish a CP-even gluino 
soft mass (${\rm arg}(M_3) = 0$) from a CP-odd gluino soft mass 
(${\rm arg}(M_3) = \pi$), through necessarily CP conserving 
processes.

Evidently several processes are affected by the gluino soft mass sign.
These include fermion masses, fermion-fermion-Higgs interactions,
trilinear scalar couplings, squark LR mixing, 
quark-squark-gaugino interactions, etc.  One-loop corrections
to the pole masses of quarks and squark LR mixing (mass)$^2$
have been calculated in e.g.\ Ref.~\cite{PBMZ}.  The one-loop
correction to the running $b$-quark mass is particularly interesting, 
since the left-right squark mixing is proportional to $\mu \tan\beta$ 
for moderate to large $\tan\beta$, and thus can give an ${\cal O}(1)$ 
correction \cite{Radiative-b-mass, COPW}.  It is straightforward to 
calculate the leading order shift in the running $b$-quark mass resulting 
from supersymmetric contributions while keeping track of the sign 
of $M_3$:\footnote{We have closely followed Ref.~\cite{COPW} for 
this calculation,
except for the differing notation for the $b$ and $t$ Yukawa couplings.}
\begin{eqnarray}
m_b &=& Y_b \, \langle H_d^0 \rangle \, (1 + \Delta m_b) 
    \label{running-b-mass-eq} \\
\Delta m_b &=& \frac{\mu \tan\beta}{16 \pi^2} \left[
    \textfrac{8}{3} g_3^2 \> {\rm sign(M_3)} M_{\tilde{g}} 
        I(m_{\tilde{b}_1}^2, m_{\tilde{b}_2}^2, M_{\tilde{g}}^2) 
  + Y_t^2 A_t I(m_{\tilde{t}_1}^2, m_{\tilde{t}_2}^2, \mu^2) \right]
  \label{delta_mb-eq}
\end{eqnarray}
where $Y_t$ and $Y_b$ are the top and bottom Yukawa couplings,
$m_{\tilde{b}_1}, m_{\tilde{b}_2}$ are the $b$-squark
masses, $m_{\tilde{t}_1}, m_{\tilde{t}_2}$ are the $t$-squark
masses, $M_{\tilde{g}}$ is the physical (real, positive definite)
gluino mass, and
\begin{eqnarray*}
I(a, b, c) &=& 
    \frac{a b \ln \frac{a}{b} + b c \ln \frac{b}{c} + a c \ln \frac{c}{a}}{
    (a - b) (b - c) (a - c)} \; .
\end{eqnarray*}
Notice that the correction to the running $b$ mass consists of a sum 
over two contributions:  one piece proportional to 
$\textfrac{8}{3} g_3^2 M_{\tilde{g}}$ (from Fig.~\ref{one-loop-fig}(a) 
with a gluino in the loop), and the other proportional to $Y_t^2 A_t$ 
(from Fig.~\ref{one-loop-fig}(a) with Higgsinos in the loop).
As we noted above, the sign of the trilinear scalar coupling in 
gauge mediation is opposite to that in anomaly mediation.
(Indeed, ${\rm sign}(M_3) \not= {\rm sign}(A_t)$ in both
anomaly mediation and gauge mediation.)
Consequently, the overall sign of 
$\Delta m_b$ is equal to ${\rm sign}(\mu) {\rm sign}(M_3)$, and is 
therefore equal to $-{\rm sign}(\mu)$ in anomaly mediation and 
$+{\rm sign}(\mu)$ in gauge mediation.  Independently determining
the sign of $\mu$ is therefore essential to interpret the
size of the correction to the running $b$ quark mass.
Other precision observables, such as $b \ra s\ph$ and $g-2$ may be
useful in this regard.\footnote{We thank J. Feng for discussions
on this point.}  

Another important difference is that in anomaly mediation $|A_t|$ is 
larger, and the top squarks are more widely separated in mass.  
The second term in Eq.~(\ref{delta_mb-eq}) is therefore different between 
the two models.  However, in our numerical calculations we found that 
the overall coefficient for the first term proportional to the gluino mass 
is always larger in both models, typically by a factor of $4$ to $6$.
(This observation was also emphasized in Ref.~\cite{COPW}.)
Consequently, if $\mu$ were the same in both models, $\Delta m_b$
would also be close in magnitude (but opposite in sign).  We can estimate 
the correction to $\Delta m_b$ by observing that
$M_{\tilde{g}}^2 I(m_{\tilde{b}_1}^2, m_{\tilde{b}_2}^2, M_{\tilde{g}}^2)
\sim 0.6$ is a good approximation throughout the parameter space
of both models, and thus
\begin{eqnarray}
\Delta m_b &\sim& (0.75) \frac{1}{6 \pi^2} \tan \beta \frac{\mu}{M_{\tilde{g}}}
    \> {\rm sign(M_3)} \label{delta_mb_approx-eq} \\
& \sim & \frac{\tan \beta}{79} \frac{\mu}{M_{\tilde{g}}} 
         \> {\rm sign(M_3)} \nonumber
\end{eqnarray}
where the $0.75$ conservatively accounts for the decrease in the correction
to $\Delta m_b$ due to the second term of Eq.~(\ref{delta_mb-eq}).
The ratio $\mu/M_{\tilde{g}}$ can be calculated from both models in 
the absence of additional contributions to the Higgs soft masses.  
($|\mu/M_{\tilde{g}}|$ is about $0.7$ in anomaly mediation and $0.5$ 
in gauge mediation.)  However, since no mechanism for the generation 
of $\mu$ was specified, there is no strongly reliable prediction of 
$\mu$ from the fundamental model parameters, particularly for gauge 
mediation.  Instead, the essential distinction between anomaly mediation 
and gauge mediation is sign of $\Delta m_b$: the one-loop running $b$-quark 
mass is larger (smaller) than the tree-level mass in anomaly mediation 
for $\mu$ negative (positive), and precisely opposite of this for 
gauge mediation.  This can be applied, for example, to the 
$h$-$b$-$\overline{b}$ coupling.  We find the effective coupling
at one-loop is
\begin{eqnarray*}
{\cal L} &=& \lambda_{hb\overline{b}} h b \overline{b} 
\end{eqnarray*}
where
\begin{eqnarray*}
\lambda_{hb\overline{b}} &=& - Y_b \sin \alpha 
    \left( 1 - \frac{\Delta m_b}{\tan \alpha \tan \beta} \right) \\
&=& \lambda_{hb\overline{b}}^{\rm tree}
    \left( 1 - \frac{\Delta m_b}{\tan \alpha \tan \beta} \right) \; .
\end{eqnarray*}
In addition, the Higgs mixing angle $\alpha$ is related to $\tan \beta$ 
at tree-level by
\begin{eqnarray*}
\tan \alpha = - \frac{1}{\tan \beta}
\end{eqnarray*}
which implies
\begin{eqnarray*}
\lambda_{hb\overline{b}} &\simeq& \lambda_{hb\overline{b}}^{\rm tree} 
    \left( 1 + \Delta m_b \right) \; .
\end{eqnarray*}
To illustrate the size of this correction, use Eq.~(\ref{delta_mb_approx-eq}) 
to obtain the coupling in the anomaly mediation model 
$\lambda_{hb\overline{b}}^{\rm AM}$ and the coupling in a gauge 
mediation model counterpart $\lambda_{hb\overline{b}}^{\rm GM}$.
All of the parameter dependence drops out except for $\tan\beta$
and $|\mu/M_{\tilde{g}}|$, in which we set the latter to be $0.6$ 
for comparison.  Then, the ratio of the couplings is simply
\begin{eqnarray*}
\frac{\lambda_{hb\overline{b}}^{\rm AM}}{\lambda_{hb\overline{b}}^{\rm GM}} 
    &\simeq& 1 - 2 \> \, {\rm sign}(\mu) \, |\Delta m_b| \\
    &\simeq& 1 - 0.015 \> \, {\rm sign}(\mu) \, \tan\beta \; .
\end{eqnarray*}
For $\mu < 0$ and $\tan \beta = (5$, $10$, $30$, $50)$, this 
ratio is $(1.08$, $1.15$, $1.45$, $1.76)$.  A high precision measurement
of this coupling (along with an independent determination of the sign 
of $\mu$) would therefore provide a useful distinguishing criteria between
these models.  The ability to measure this coupling at the LEP collider, 
the Fermilab Tevatron, and the LHC has been studied in 
Ref.~\cite{b-Yukawa-pheno}, although it is likely that 
a NLC or a muon collider would be necessary to approach the 
needed precision.

Another class of processes, shown in Fig.~\ref{one-loop-fig}(c), are 
the one-loop gluino corrections to the quark-squark-gaugino vertex.
These corrections arise in gaugino decay 
$\tilde{G} \ra q \tilde{q}^{(')}$ and/or squark decay 
$\tilde{q} \ra \tilde{G} q^{(')}$, depending on the kinematics.
Here $\tilde{G}$ can be any gaugino, and the prime denotes
the SU(2) doublet partner field (for decays involving a chargino).
These radiative corrections have the advantage that they
are proportional to $g_3^2$, but the disadvantage that
they are suppressed by $|M_3|^2$ and must involve squarks
that are typically much heavier than sleptons.

Several other processes involving the diagrams of Fig.~\ref{one-loop-fig} 
might be useful.  Determining the best experimental observable
depends on the precision to which particular sparticle properties 
are measured and the collider that is used.

\subsection{NLSP decay}
\label{NLSP-decay-sec}

One apparently obvious distinction between anomaly mediation and 
gauge mediation is that the Wino NLSP of a gauge-mediated model 
decays into a gravitino plus a photon.  Even if the decay length 
is significantly larger than the scale of a collider detector, 
a statistically significant excess of sparticle production events 
with one (or two) hard photon(s) could experimentally establish that 
the Wino is unstable (otherwise, place a lower bound on its decay length).  
To confirm that the Wino is unstable, in principle only a few events 
would be necessary, as long as SM backgrounds can be reduced to a 
negligible level.  The probability that one NLSP decays within the
distance $L$ ($\ll L_{\rm NLSP}$) of order the detector size,
where $L_{\rm NLSP} = c / \Gamma_{\rm NLSP}$, 
is $1 - e^{- L/L_{\rm NLSP}} \sim L/L_{\rm NLSP}$.
Thus, one would expect about one NLSP decay for every 
$L_{\rm NLSP}/L$ NLSPs produced.
However, as discussed above, the decay length could be 
enormous compared with the scale of a detector, and thus 
be well beyond the range of ordinary collider experiments.
In this case, the gauge-mediated Wino-related collider signals are 
indistinguishable from anomaly mediation (or any other supersymmetric 
model that predicts a stable Wino).  The difference might be detectable
with a very long baseline experiment to measure NLSP decay, or, if there
is a cosmologically significant relic density of Wino LSPs 
of anomaly mediation \cite{GGW, MR}, using dark matter detection 
experiments.

\section{Higher order corrections}
\label{higher-order-sec}

The equivalence of the gauge-mediated and anomaly-mediated gaugino masses
holds to leading order (LO), but not to higher orders.  This is simply
a consequence of the fundamentally different origin of the soft masses.
Higher order corrections are generally expected to be suppressed
by a one-loop factor $1/16 \pi^2$ times order one couplings and
coefficients, relative to the leading order corrections. 
There are, however, important next-to-leading (NLO) 
two-loop corrections to the gaugino 
masses \cite{two-loop} that are much larger than might be naively
expected \cite{Kribs}.  The most important correction relevant
to this discussion is the two-loop contribution to $M_2$ due
to the $g_2^2 g_3^2 M_3$ term.  In anomaly mediation, this
takes the form
\begin{eqnarray}
M_2^{\rm AM} |_{\rm NLO} &=& M_2^{\rm AM} |_{\rm LO} 
    \left[ 1 + \frac{B_{23}^{(2)} g_3^2}{B_2^{(1)} 16 \pi^2} \right]
\end{eqnarray}
where $B_{23}^{(2)} = 24$ is the rather large two-loop coefficient.  
The NLO result is about $20\%$ larger than the LO result 
at the weak scale \cite{GGW}.  In gauge mediation, there is also
a correction from the renormalization group evolution that explicitly
depends on the logarithm of the ratio of scales.  Specifically,
this correction can be approximated as
\begin{eqnarray}
M_2^{\rm GM} |_{\rm NLO} &=& M_2^{\rm GM} |_{\rm LO} 
    \left[ 1 - \frac{n_3 B_{23}^{(2)} g_3^4}{n_2 (16 \pi^2)^2} 
    \ln \frac{\Mmess}{M_2} \right] \; .
\end{eqnarray}
The gauge-mediated NLO expression for $M_2$ is typically a 
few percent \emph{smaller} than the LO result, depending on
the messenger scale $\Mmess$.  

The other gaugino masses $M_1$ and $M_3$ receive at most a few 
percent correction to their LO values from the NLO pieces of the 
$\beta$-function, in both AM and GM models.  Since one-loop threshold 
corrections are of the same order (if not significantly larger, 
especially for $M_3$), the NLO $\beta$-function corrections 
for these masses can be neglected.

The large correction to $M_2$ in the anomaly-mediated approach naively 
suggests that accurately measuring the ratio $M_1/M_2$ would distinguish
anomaly mediation from gauge mediation at next-to-leading order.  
However, it is not hard to imagine that messengers could generate 
the approximate proportion ($n_1$ : $n_2$ : $n_3$) $\sim$ ($33/5$ : $1.2$ : 
$3$).  Perhaps the simplest possibility is to assume that 
$F/M^2$ for the $L + \overline{L}$ multiplet is about $0.9$, 
while all of the other multiplets have $F/M^2 \ll 1$.  This
generates a rather large positive one-loop threshold correction 
of about the right size for $M_2$ only.\footnote{In this case, 
higher order corrections from messenger contributions could also 
be important \cite{Strumia-AHGLR}.}
The multi-singlet model with differing $F$-terms could also reproduce
this ratio, but requires three different singlets coupling to the 
three SM components of the $10 + \overline{10}$, with $F$-terms in 
the proportion ($F_{X_Q}/M_{X_Q}$ : $F_{X_u}/M_{X_u}$ : $F_{X_e}/M_{X_e}$) 
$\sim$ ($4$, $22$, $25$).
Thus, the gauge-mediated Wino NLSP models could approximately 
reproduce the NLO gaugino mass predictions of anomaly mediation, 
although one must make some slight modification to the messenger 
sector that is admittedly rather ad hoc.

\section{Conclusions}
\label{conclusions-sec}

There exist gauge-mediated Wino NLSP models that predict the gaugino
masses, the first and second generation squark masses, 
and potentially the left-handed first and second generation
slepton masses are nearly equivalent to the predictions of an
anomaly-mediated model with a small universal additional scalar mass.
The sbottom masses, stau masses, heavier Higgs masses,
and heavier chargino and neutralino masses are in general 
somewhat different due to a differing value of $\mu$ (determined
from EWSB constraints).  But, we have argued that these observables are 
not useful distinctions between the two classes of models because 
the mechanism for the dynamical generation of $\mu$ in gauge mediation, 
which could affect the Higgs soft masses, is unknown.
This leaves only the ratios $(m_{\ell_R}^2 - m_{\ell_L}^2)/m_{\ell_L}^2$ 
and $(m_{\tilde{t}_2}^2 - m_{\tilde{t}_1}^2)/m_{\tilde{t}_2}^2$ as
useful parameter-independent experimental criteria to distinguish 
between anomaly mediation and the gauge-mediated models discussed
here.  These scalar mass ratios remain useful distinguishing criteria 
even if the universality of the additional scalar mass $m_0$ is relaxed,
but only if the left-handed and right-handed additional mass 
contribution is the same.

The differing sign of the gluino soft mass is perhaps the clearest
distinction between anomaly mediation and gauge mediation,
and does not depend on the universality of $m_0$.
Several one-loop processes are sensitive to this sign.  Probably
the most promising way to determine the sign of the gluino soft 
mass is to accurately measure the $h$-$b$-$\overline{b}$ coupling
and compare it with the tree-level expectation.  Given an
independent measurement of the sign of $\mu$, this coupling
is significantly enhanced (diminished) in anomaly-mediation
(gauge-mediation) with $\mu < 0$, and vice versa for $\mu > 0$.
The precise size of the correction is strongly dependent on
$\tan \beta$ and $\mu/M_{\tilde{g}}$, with a weaker dependence 
on other MSSM quantities.

The NLSP of any gauge-mediated model is unstable.  Observing
the decay of NLSP $\ra$ gravitino plus a photon would strongly
suggest gauge mediation, although the decay length could
be well beyond the sensitivity of any collider experiment.
In this scenario, a long-lived Wino NLSP of gauge mediation
is indistinguishable from a stable Wino LSP of anomaly mediation
in ordinary collider experiments.  Establishing the (in)stability
of the Wino would thus require either a very long baseline experiment 
to search for its decay, or a dark matter detection experiment to search 
for a cosmologically significant relic density of Wino LSPs.

\section*{Acknowledgments}
\indent

I thank L.~Everett, J.~Feng, T.~Gherghetta, D.R.T.~Jones, 
G.L.~Kane, D.E.~Kaplan, S.P.~Martin, R.~Rattazzi, and C.~Wagner 
for helpful discussions and correspondence.  
I also thank the Aspen Center for Physics for hospitality 
where part of this work was completed.  This work was supported in part 
by the U.S. Department of Energy under grant number DOE-ER-40682-143.

\newpage

\begin{appendix}
\refstepcounter{section}

\section*{Appendix~\thesection:~~Comparison of example models' spectra}
\label{example-models-app}

In Table~\ref{example-models-table} four examples of models
exhibiting the characteristics discussed in Sec.~\ref{distinctions-sec}
are given.  For all of the models, one-loop $\beta$-functions
are used, and the masses of the sparticles are given at tree-level 
(i.e., one-loop corrections resulting from the conversion from an 
$\overline{\rm MS}$ mass to a pole mass were not included).  
In all cases the same value of $\tan\beta$, ${\rm sign}(\mu)$, and 
$m_{3/2}^{\rm AM} = F/M$ were used.  Examples (a) and (b) are anomaly 
mediation models with slightly differing values of $m_0$.  Examples (c) 
and (d) are gauge mediation models with differing values of the Higgs 
soft masses.  In example (c) only the gauge-mediated contribution to the
Higgs soft masses were included, whereas in example (d) additional
contributions to the Higgs soft masses were included such that
the resulting Higgs soft masses at the weak scale were the same
as in the anomaly mediation model example (a).  The latter has
the effect of generating a $\mu$ in gauge mediation that is the 
same as anomaly mediation.

Clearly $M_1$, $M_2$, $|M_3|$ and $m_h$ are identical, and the heavy 
first (and second) generation squark masses are the same to within 
about 2\%.  If $\mu$ is the same in both anomaly mediation and 
gauge mediation, then all the gaugino masses and the heavy Higgs masses 
are also the same to within 2\%.  The differences between the models 
arise in the third generation and the slepton masses.  Again, to within 
a few percent, example (a) and examples (c),(d) have the same left-handed 
first (and second) generation slepton masses.  Similarly example (b) and 
examples (c),(d) have the same right-handed first (and second)
generation slepton mass.  (The additional universal scalar
mass $m_0$ in anomaly mediation was chosen to give these results.)
This illustrates the important general observation that, all other
things equal, it is \emph{not} possible to find the same left-handed 
and right-handed slepton masses in an anomaly mediation and a gauge 
mediation model, assuming the universality of $m_0$.  

In general the stau masses can be quite different, unless
both the right-handed slepton masses and $\mu$ are the same 
[compare examples (b) and (d)].  Again, however, the ratio of
the right-handed first (or second) generation slepton mass
with the left-handed first (or second) generation slepton mass
remains significantly different.  The lightest $b$ squark mass
is lighter in the anomaly mediation models by about 7\%
(while the heavier $b$ squark mass is comparable across all example
models).  The lighter (heavier) $t$-squark mass is lighter in the 
anomaly mediation models by about 14\% (6\%).  Finally, the
correction to the running $b$ mass, Eq.~(\ref{delta_mb-eq}), 
is close in magnitude when comparing examples (a),(b) with (d), 
but opposite in sign between anomaly mediation and gauge mediation.  
The magnitude of the correction is smaller for example (c) due to 
the smaller value of $\mu$.

\begin{table}
\begin{center}
\begin{tabular}{c|cccc} \hline\hline
     & \multicolumn{2}{c}{Anomaly mediation} & 
       \multicolumn{2}{c}{Gauge mediation} \\
     & Example (a) & Example (b) & Example (c) & Example (d) \\ \hline\hline
$m_{3/2}^{\rm AM}$ & $5 \times 10^4$ & $5 \times 10^4$ & & \\
$m_0$ & $291$ & $327$ & & \\
$F/M$ & & & $5 \times 10^4$ & $5 \times 10^4$ \\
$\Lambda_{DSB}$ & & & $10^7$ & $10^7$ \\
$\tan\beta$ & $20$ & $20$ & $20$ & $20$ \\
${\rm sign}(\mu)$ & $+$ & $+$ & $+$ & $+$ \\ \hline
$\mu$ & $810$ & $811$ & $553$ & $813$ \\
$M_1$ & $454$ & $454$ & $454$ & $454$ \\
$M_2$ & $133$ & $133$ & $133$ & $133$ \\
$M_3$ & $-1061$ & $-1061$ & $1061$ & $1061$ \\
$m_{\tilde{C}_1}$, $m_{\tilde{C}_2}$ 
      & $131$, $819$ & $131$, $819$ & $129$, $565$ & $131$, $821$ \\
$m_{\tilde{N}_1}$, $m_{\tilde{N}_2}$ 
      & $131$, $451$ & $131$, $451$ & $129$, $445$ & $131$, $451$ \\
$m_{\tilde{N}_3}$, $m_{\tilde{N}_4}$ 
      & $814$, $818$ & $815$, $819$ & $558$, $570$ & $817$, $821$ \\
$m_{\tilde{t}_1}$, $m_{\tilde{t}_2}$
      & $770$, $935$ & $774$, $941$ & $895$, $993$ & $894$, $994$ \\
$m_{\tilde{b}_1}$, $m_{\tilde{b}_2}$
      & $882$, $988$ & $890$, $998$ & $952$, $1001$ & $946$, $1007$ \\
$m_{\tilde{\tau}_1}$, $m_{\tilde{\tau}_2}$
      & $111$, $260$ & $182$, $298$ & $202$, $293$ & $182$, $306$ \\
$m_{\tilde{u}_L}$, $m_{\tilde{u}_R}$
      & $1029$, $1034$ & $1039$, $1044$ & $1017$, $1017$ & $1017$, $1017$ \\
$m_{\tilde{e}_L}$, $m_{\tilde{e}_R}$
      & $229$, $224$ & $273$, $269$ & $229$, $276$ & $229$, $276$ \\
$m_h$, $m_H$ 
      & $121$, $710$ & $121$, $724$ & $121$, $553$ & $121$, $712$ \\
$\Delta m_b$ 
      & $-0.17$ & $-0.17$ & $0.12$ & $0.18$ \\ \hline\hline
\end{tabular}
\end{center}
\caption{Example models are shown with the input parameters
and some of the masses of the resulting weak scale spectrum
using one-loop renormalization group evolution.
Examples (a) and (b) are anomaly mediation models with slightly
differing values of $m_0$.  Examples (c) and (d) are
gauge mediation models with messenger content defined by
Eq.~(\ref{mess-content-eq}), and (c) no additional contributions 
to the Higgs soft masses, or (d) additional contributions such 
that $m_{H_u}^2$ and $m_{H_d}^2$ are the same as those in the anomaly 
mediation example (a).  All quantities have dimensions of GeV except 
of course for $\tan \beta$, ${\rm sign}(\mu)$, and $\Delta m_b$, which 
are dimensionless.}
\label{example-models-table}
\end{table}

\refstepcounter{section}

\newpage

\section*{Appendix~\thesection:~~The Lagrangian for a complex gluino soft mass}
\label{gluino-app}

Assume the gluino soft mass is complex $M_3 = |M_3| e^{i \theta}$ and
that the phase is physical (i.e., cannot be rotated into some other 
soft breaking quantity).  The relevant pieces of the softly broken 
interaction Lagrangian involving the gluino are
\begin{eqnarray*}
{\cal L}_{\tilde{g}} &=& {\cal L}_{\tilde{g}, {\rm gauge}} 
                            + {\cal L}_{\tilde{g}, {\rm chiral}}
\end{eqnarray*}
where
\begin{eqnarray*}
{\cal L}_{\tilde{g}, {\rm gauge}} &=& 
    - i \lambda^{\dagger a} \overline{\sigma}^\mu {\cal D}_\mu^a 
    \lambda^{a} - \textfrac{1}{2} (M_3 \lambda^a \lambda^a + h.c.) \\
{\cal L}_{\tilde{g}, {\rm chiral}} &=& 
    \sqrt{2} g_3 \sum_i \left( i \phi_i^* T_i^a \psi_i \lambda^a + h.c. 
    \right) \; .
\end{eqnarray*}
$\lambda^a$ is the 2-component gluino spinor with $a=1 \ldots 8$, 
the $\sum_i$ is over all chiral multiplets, $T_i^a$ is the SU(3) matrix 
for the $i^{\rm th}$ representation, $\phi$ and $\psi$ are the complex
scalar and 2-component spinor for the chiral multiplet, and the
covariant derivative acting on the gluino is
\begin{eqnarray*}
{\cal D}_\mu^a \lambda^a &=& \partial_\mu \lambda^a 
                                - g_3 f^{abc} A_\mu^b \lambda^c \; .
\end{eqnarray*}
Under the field rotation $\lambda^a \ra \lambda^a e^{-i \theta/2}$,
${\cal L}_{\tilde{g}, {\rm gauge}}$ is invariant (except, 
of course, that $M_3 \ra |M_3|$), while 
${\cal L}_{\tilde{g}, {\rm chiral}}$ becomes
\begin{eqnarray*}
  \sqrt{2} g_3 \sum_i \left( i \phi_i^* T_i^a \psi_i \lambda^a e^{-i \theta/2} 
                               + h.c. \right) \; .
\end{eqnarray*}
In four component notation, the quark--squark--gluino interaction 
Lagrangian (see Eq.~(C89) in Ref.~\cite{HaberKane}) in the MSSM becomes
\begin{eqnarray}
   - \sqrt{2} g_3 T^a_{jk} \sum_i \left( 
   \tilde{q}_{iL}^{j*} \overline{\tilde{g}}_a P_L q_i^k e^{-i \theta/2}
   - \tilde{q}_{iR}^{j*} \overline{\tilde{g}}_a P_R q_i^k e^{i \theta/2}
   + h.c. \right)
\label{quark-squark-gluino-lagrangian-eq}
\end{eqnarray}
where the sum is over all quarks $i=u,d,c,s,t,b$.  This agrees with
the result found in Ref.~\cite{IN98}.

\end{appendix}

\end{document}